\documentclass[doublecol,linenumbers]{epl2} 

\title{Coupled harmonic oscillator in a multimode harmonic oscillator bath: Derivation of quantum propagator and master equation using white noise analysis}
\shorttitle{White noise path integration of coupled oscillators in a single multimode oscillator bath} 

\author{B. M. Butanas Jr\inst{1,2} \and R. C. F. Caballar\inst{1} }
\shortauthor{B. M. Butanas Jr \etal}

\institute{                    
  \inst{1} Theoretical Physics Group, National Institute of Physics, University of the Philippines, Diliman Quezon City, 1101 Philippines\\
  \inst{2} Department of Physics, Central Mindanao University, University Town, Musuan, Maramag, Bukidnon, 8710 Philippines
}
\pacs{03.65.Yz}{Decoherence; open systems; quantum statistical methods}
\pacs{02.50.Ey}{Stochastic processes}
\pacs{02.50.Fz}{Stochastic analysis}

\abstract{This paper presents an application of white noise functional approach to derive the quantum propagator and the evolution of the reduced density matrix for an open quantum system consisting of coupled harmonic oscillators which are coupled to a bath consisting of a multimode harmonic oscillator. It is shown that the full quantum propagator is a product of three individual propagators. These propagators were obtained after two successive transformations of the coordinates for the system, and the coordinates for the coupling between the bath and one of the two harmonic oscillators in the system. The obtained propagator is then used to derive the time evolution equation for the density matrix describing the system. The method used to derive the propagator of the open quantum system considered in this paper shows promise for analyzing open quantum systems, due to its ease of use and mathematical rigor.} 

\begin{document}

\maketitle

\section{Introduction}
Interactions between a quantum system and its surrounding environment have been a subject of intense study for the past decades \cite{i.a}-\cite{i.d}. These studies have enabled us to understand better the mechanism behind various quantum processes such as decoherence \cite{i.e}-\cite{i.n}. An example of such a mechanism is the Caldeira-Leggett (CL) model \cite{i.a, a.a}. It is a microscopic quantum system-bath prototype that enables us to describe dissipation phenomena in solid state physics, quantum tunneling and quantum computing\cite{a.a1, a.a1a}. The CL model describes a quantum system with an arbitrary potential interacting with an environment modeled as an infinite number of harmonic oscillators \cite{i.a}. The influence-functional method of Feynman and Vernon \cite{a.a2} was used to analyze this model. In doing so, the master equation describing the dissipative dynamics of the system was obtained. 

The exact master equation for the CL model with the system composed of a single harmonic oscillator interacting with an environment of infinite harmonic oscillators was solved using the influence functional \cite{i.b}, Wigner function \cite{a.a3} and quantum trajectories method \cite{a.a4}. The propagator for this particular CL model was solved in refs. \cite{i.e, a.b}. These studies can be extended and generalized towards the analysis of a system composed of N harmonic oscillators coupled to an environment modeled as an ensemble of harmonic oscillators. Such an extension and generalization is vital in understanding macroscopic quantum phenomena such as decoherence since any quantum system, and the environment with which it is interacting, can be decomposed into a number of components which are modeled as harmonic oscillators.  

In this paper, we consider a system of coupled harmonic oscillators interacting with an environment which is modeled as a single multimode harmonic oscillator. Furtheremore, as compared with refs. \cite{i.b, i.e, a.a, a.b}, this method presents a new way of deriving the quantum propagator and master equation using the method of white noise analysis invented by Hida \cite{a.c}. As compared to the influence functional method which is considered to be mathematically ill-defined due to the presence of the Lebesgue measure, white noise analysis is a mathematically well-defined method, and is a powerful tool in evaluating the Feynman path integral \cite{a.d}. It is our motivation here to show the promise of white noise analysis in evaluating propagators for an open quantum systems.     

This paper is organized as follows. The first section will present the system and the bath, together with their corresponding Hamiltonians, considered in this study. The second and third sections will discuss some basics on white noise analysis and the recasting of Feynman path integral in the context of white noise analysis following ref. \cite{a.d}. Then, the fourth section tackles the derivation of the master equation from the propagator. Finally, the fifth section will present the application of white noise analysis to derive the quantum propagator and the master equation, before we conclude with a short summary.

\section{Coupled Harmonic Oscillators in a Bath}
\label{coupled}

The Hamiltonian $H_{S}$ of the coupled harmonic oscillator system immersed in a bath of harmonic oscillators with Hamiltonian $H_{B}$ is defined, respectively, as \cite{a.a} 
\begin{eqnarray}
H_{S} = \frac{p_{1}^2}{2m} + \frac{1}{2}m\omega^2 x_{1}^2 + \frac{p_{2}^2}{2m} + \frac{1}{2}m\omega^2 x_{2}^2 + \lambda x_{1}x_{2}, 
\label{coupled hs}
\\
H_{B} = \frac{p_{q}^2}{2m} + \frac{1}{2}m\omega_{q}^2 q^2, \hspace{0.5cm} H_{SB} =  Cq(x_{1}+x_{2}),
\label{hb}
\end{eqnarray}
where $x, q$, $p$, and $\omega$, are the corresponding positions, momenta, and frequencies of the system and bath oscillators, while $\lambda$ is the coupling constant of the system-system interaction. Moreover, we assume that the bath coordinate is linearly coupled to the system with Hamiltonian $H_{SB}$ given above with coupling constant $C$. Thus, the total Hamiltonian can be written as 
\begin{equation}
H = H_{S} + H_{B} + H_{SB}.
\label{totalH}
\end{equation}

\section{White Noise Analysis Fundamentals}
\label{whitenoise}

Formally, a stochastic process like Brownian motion obeys the stochastic differential equation given by
\begin{equation}
dX = a(t,X) dt + b(t,X) dB(t),
\label{stochasticdifeqn}
\end{equation}
where $X$ describes the Brownian motion, $B(t)$ is the Wiener process, $a(t,X)$ and $[b(t,X)]^2$ are the drift and diffusion coefficients, respectively. Then, one can solve for the corresponding Langevin equation, which is
\begin{equation}
\dot{X} = a(t,X) + b(t,X) \omega(t),
\label{langevin}
\end{equation}
where $\dot{X}=\frac{dX}{dt}$ and $\omega(t)=\frac{dB(t)}{dt}$, is interpreted as the velocity of Brownian motion, and is called the Gaussian white noise. Furthermore, we can rewrite the Gaussian white noise in terms of Wiener's Brownian motion i.e. $B(t)=\int_{t_{o}}^{t}\omega(\tau) d\tau=\left\langle \omega, 1_{[t_{o},t)}\right\rangle$, where we define $\left\langle \omega, \xi \right\rangle \equiv \int_{t_{o}}^{t}\omega(\tau) \xi(\tau) d\tau$.

Now, a key feature of Hida's formulation is the treatment of the set $\omega(\tau)$ at different instants of time, $\left\{ \omega(\tau); t \in \Re \right\}$ as a continuum coordinate system. For the sum over all routes or histories in the path integral, paths starting from initial point $x_{o}$ and propagating in Brownian fluctuations are parametrized within the white noise framework as
\begin{equation}
x(t) = x_{o} + \int_{t_{o}}^{t}\omega(s) ds.
\label{parametizewithmemory}
\end{equation}
Eq. (\ref{parametizewithmemory}) shows how the value of $x(t)$ is affected by its history, or earlier values of the modulated white noise variable $\omega(s)$ as $s$ ranges from $t_{o}$ to $t$. 

A further key feature of white noise analysis is that it operates in the Gelfand triple \cite{b.a} $S \subset L^{2} \subset S^{*}$, linking the spaces of a Hida distribution $S^{*}$ and test function $S$ through a Hilbert space of square integrable functions $L^{2}$. Using Minlos' theorem we can formulate a Hida white noise space $(S^{*},B,\mu)$ where $\mu$ is the probability measure and $B$ is the $\sigma$-algebra generated on $S$, and define a characteristic functional $C(\xi)$ given by
\begin{equation}
C(\xi) = \int_{S^{*}}\exp \left( \left\langle \omega, \xi \right\rangle \right) d\mu(\omega) = \exp \left( -\frac{1}{2}\int \xi^2 d\tau \right),
\label{characteristic}
\end{equation}
where $\xi \in S$ and the white noise Gaussian measure $d\mu(\omega)$ is given by
\begin{equation}
d\mu(\omega) = N_{\omega}\exp \left( -\frac{1}{2}\int \omega^2(\tau) d\tau \right) d^{\infty}\omega,
\label{gaussianmeasure}
\end{equation}
with $N_{\omega}$ as a normalization constant. The exponential term in $d\mu(\omega)$ is responsible for the Gaussian fall-off of the propagator function.

Now, the evaluation of the Feynman integral in the context of white noise analysis is carried out by the evaluation of the Gaussian white noise measure $d\mu(\omega)$. There are two important Gaussian white noise measure evaluation methods; these are the use of $T$- and $S$-transforms. For the $T$-transform of a generalized white noise functional $\Phi(\omega)$ we have the form
\begin{equation}
T\Phi(\xi) = \int_{S^{*}}\exp \left( i\left\langle \omega, \xi \right\rangle \right) \Phi(\omega) d\mu(\omega),
\label{Ttransform}
\end{equation}
similar to that of an infinite-dimensional Gauss-Fourier transform. On the other hand, the $S$-transform is related to the $T$-transform as follows:
\begin{equation}
S\Phi(\xi) = C(\xi) T\Phi(-i\xi),
\label{Stransform}
\end{equation}
where $C(\xi)$ is the characteristic functional given in eq. (\ref{characteristic}).

\section{Feynman Quantum Propagator as a White Noise Functional}
\label{feymanevaluation}

The propagator for the quantum mechanical oscillator has the following form as derived by Feynman \cite{c.a}:
\begin{equation}
K(x,x_{o};\tau) = \int \exp \left( \frac{i}{\hbar}S \right) D[x],
\label{propa2}
\end{equation}
where $S$ is the classical action defined as $S=\int L ~dt$ with $L$ as the Lagrangian of the system, and $D[x]$ is the infinite-dimensional Lebesgue measure. Eq. (\ref{propa2}) sums over all the possible paths taken by a system/particle from an initial point $x(t_{o}=0) = x_{o}$ to a final point $x(t) = x$. Now, to rewrite this using the white noise analysis approach \cite{b.a, a.d} we introduce the parametrization of the path given by
\begin{equation}
x(t) = x_{o} + \sqrt{\frac{\hbar}{m}}\int_{0}^{t}\omega(\tau)d\tau,
\label{parametrize}
\end{equation}
Then taking the derivative of eq. (\ref{parametrize}), substituting it into eq. (\ref{exponential}) and simplifying the resulting equation, we obtain the exponential expression in the right hand side of eq. (\ref{propa2}) as
\begin{equation}
\exp\left(\frac{i}{\hbar}S\right)= \exp \left[ \frac{i}{2}\int_{0}^{t}\omega(\tau)^2 d\tau \right] \exp \left[ -\frac{i}{\hbar}\int_{0}^{\tau}V(x) d\tau \right].
\label{exponential}
\end{equation} 
On the other hand, evaluation of the Lebesgue measure $D[x]$ leads to an integration over the Gaussian white noise measure $d\mu(\omega)$ in the relation
\begin{equation}
D[x]=\lim_{N\rightarrow \infty}\prod^{N}(A_{j})\prod^{N-1}(dx_{j})=Nd^{\infty}x, 
\label{gaussianmeasure0}
\end{equation} 
with
\begin{equation}
Nd^{\infty}x \rightarrow Nd^{\infty}\omega = \exp \left[ \frac{1}{2}\int_{0}^{t}\omega(\tau)^2 d\tau \right] d\mu(\omega),
\label{gaussianmeasure}
\end{equation}
where $N$ is the normalization constant. However, the path parametrization of the Brownian motion in eq. (\ref{parametrize}) shows that only the initial point $x_{o}$ is fixed while the final point is fluctuating. Thus, to fix the endpoint we use the Fourier decomposition of a Donsker delta function, $\delta(x(t)-x)$, defined as
\begin{equation}
\delta(x(t)-x)=\frac{1}{2\pi}\int_{-\infty}^{+\infty}\exp \left( i\lambda (x(t)-x) \right) d\lambda,
\label{donsker}
\end{equation}
such that at time $t$ the particle is located at $x$. Finally, with eqs. (\ref{exponential}), (\ref{gaussianmeasure}) and (\ref{donsker}) we now write the Feynman propagator in the context of white noise analysis as
\begin{eqnarray}
K(x,x_{o};\tau)&=&N \int \exp \left[ \frac{i+1}{2}\int_{0}^{t}\omega(\tau)^2 d\tau \right] 
\label{whitenoisepropagator}
\\
&\times&\exp \left[ -\frac{i}{\hbar}\int_{0}^{\tau}V(x)d\tau \right] \delta(x(t)-x) d\mu(\omega).
\nonumber
\end{eqnarray}

\section{The Quantum Master Equation}
\label{master}

To obtain the master equation, we start from the evolution of the density matrix given by
\begin{equation}
\frac{d}{dt}\rho (t) = -\frac{i}{\hbar} [H(t),\rho (t)] = \textbf{L(t)}\rho(t),
\label{von}
\end{equation}
known as the \textit{Liouville-von Neumann} equation where $\rho (t)$, $H(t)$ and $\textbf{L(t)}$ are the density matrix, Hamiltonian of the total system and Liouville \textit{super-operator}, respectively. Formally, eq. (\ref{von}) can be written as
\begin{equation}
\rho (t) = T_{\leftarrow} \exp \left[ -\frac{i}{\hbar}\int_{0}^{\tau}\textbf{L(t)} dt \right] \rho (t_{o}),
\label{vonformal}
\end{equation}
where $T_{\leftarrow}$ describes the usual chronological time-ordering operator. Moreover, since we want to describe the dynamics of the relevant system we take the trace of the bath leading to the reduced density matrix $\rho_{S}(t)=Tr_{B}(\rho(t))$. This gives us information on the dynamics of the system while being influenced by the bath \cite{i.c}. Then, we can rewrite the evolution of the density matrix in eq. (\ref{vonformal}) as 
\begin{equation}
\rho_{S}(t) = J(t,t_{o}) \rho (t_{o}),
\label{vonformal2}
\end{equation}
where $J(t,t_{o})$ is the Liouville space propagator or simply the evolution operator given by
\begin{equation}
J(x,x_{o};x',x'_{o};\tau) \equiv \left\langle T_{\leftarrow} \exp \left[ -\frac{i}{\hbar}\int_{0}^{\tau}\textbf{L(t)} dt \right] \right\rangle_{B},
\label{reducedpropa}
\end{equation}
with the symbol $\left\langle \cdots \right\rangle_{B}$ corresponds to the $Tr_{B}(\cdots \rho_{B}(t))$ \cite{i.c}. Now in the context of white noise path integral formalism, by correspondence, we get the relation $J(x,x_{o};x',x'_{o};\tau)=K(x,x_{o};\tau)K^{*}(x',x'_{o};\tau)$ where the propagator $K(x,x_{o};\tau)$ is given by eq. (\ref{whitenoisepropagator}).

\section{Evaluation of the Feynman Path Integral for the Coupled Harmonic Oscillators in a Bath using White Noise Analysis}
\label{evaluation}

With the total Hamiltonian given in eq. (\ref{totalH}), we solve for the total Lagrangian using Hamilton's canonical transformations 
\begin{equation}
\dot{q}_{k} = \frac{\partial H}{\partial p_{k}}, \hspace{0.5cm} -p_{k} = \frac{\partial H}{\partial q_{k}},
\nonumber
\end{equation}
and the relation given by 
\begin{equation}
L = \sum_{k}p_{k}\dot{q}_{k} - H,
\nonumber
\end{equation}
which yields the total Lagrangian as $L = L_{1} + L_{2}$ where
\begin{eqnarray}
{L_{1}}&=&\frac{1}{2}m\dot{x}_{1}^2 - \frac{1}{2}m\omega^2{x}_{1}^2 + \frac{1}{2}m\dot{x}_{2}^2 - \frac{1}{2}m\omega^2{x}_{2}^2 
\nonumber
\\
& & - \lambda x_{1}x_{2} - Cq(x_{1}+x_{2}),
\label{l1}
\\ 
{L_{2}}&=&\frac{1}{2}m\dot{q}^2 - \frac{1}{2}m\omega_{q}^2{q}^2.
\label{l2}
\end{eqnarray}
Then we utilize a transformation \cite{c.c} which we can use to decouple ${x}_{1}$ and ${x}_{2}$, and gives the relation
\begin{eqnarray}
x_{1} &=& q_{1} \cos \phi + q_{2} \sin \phi,
\label{x1}
\\
x_{2} &=& -q_{1} \sin \phi + q_{2} \cos \phi.
\label{x2}
\end{eqnarray}
It can be shown, by differentiating eqs. (\ref{x1}) and (\ref{x2}) and substituting them into eq. (\ref{l1}), that
\begin{eqnarray}
L_{1}&=&\frac{1}{2}m\dot{q}_{1}^2 - \frac{1}{2}m\omega^2{q}_{1}^2 + \frac{1}{2}m\dot{q}_{2}^2 - \frac{1}{2}m\omega^2{q}_{2}^2 
\label{l1transformed}
\\
&-&\lambda {q}_{1}^2 \cos \phi \sin \phi - \lambda {q}_{1}{q}_{2}\cos 2\phi - \lambda {q}_{2}^2 \cos \phi \sin \phi
\nonumber
\\
&-&Cqq_{1}(\cos \phi - \sin \phi) - Cqq_{2}(\cos \phi + \sin \phi).
\nonumber
\end{eqnarray}  
Further regrouping and simplification yields
\begin{eqnarray}
L_{1}&=&\frac{1}{2}m\dot{q}_{1}^2 + \frac{1}{2}m\dot{q}_{2}^2 + \alpha {q}_{1}^2 + \beta{q}_{2}^2  
\nonumber
\\
&-&\gamma{q}_{1}{q}_{2} - Cq \mu q_{1} + Cq \nu q_{1},
\label{l1transformedsimp}
\end{eqnarray}
where
\begin{eqnarray}
\alpha&=&\frac{1}{2}m\omega^2 + \lambda \cos \phi \sin \phi,
\label{alpha}
\\
\beta&=&\frac{1}{2}m\omega^2 - \lambda \cos \phi \sin \phi,
\label{beta}
\\
\gamma&=&\lambda \cos 2\phi,
\label{gamma}
\\
\mu&=&\cos \phi - \sin \phi,
\label{mu}
\\
\nu&=&\cos \phi + \sin \phi.
\label{nu}
\end{eqnarray}
In order to eliminate the system-system coupling, $\gamma$ must vanish, so that eq. (\ref{gamma}) yields
\begin{equation}
\phi = \frac{(2n+1)\pi}{4},
\label{gammasolution}
\end{equation}
where $n=0,1,2,...$. Now when the condition of eq. (\ref{gammasolution}) is imposed, eqs. (\ref{alpha}),(\ref{beta}),(\ref{mu}) and (\ref{nu}) can be simplified and hence we can rewrite the total Lagrangian as $L=L_{1}+L_{2}+L_{3}$ where
\begin{eqnarray}
L_{1} &=& \frac{1}{2}m\dot{q}_{1}^2 + \frac{1}{2}m\Omega_{1}^2 q_{1}^2,
\label{l1new}
\\
L_{2} &=& \frac{1}{2}m\dot{q}_{2}^2 + \frac{1}{2}m\Omega_{2}^2 q_{2}^2 - \sqrt{2}Cqq_{2},
\label{l2new}
\\
L_{3} &=& \frac{1}{2}m\dot{q}^2 + \frac{1}{2}m\omega_{q}^2{q}^2,
\label{l3new}
\end{eqnarray}
with newly defined frequencies $\Omega_{1}^2=\omega^2 + \frac{\lambda}{m}$ and $\Omega_{2}^2=\omega^2 - \frac{\lambda}{m}$. Notice that we have successfully decoupled the two coupled oscillators in the system, as can be seen in eqs. (\ref{l1new})-(\ref{l3new}). On the other hand, we are now left with a coupling between the system and the bath coordinates $q_{2}$ and $q$ respectively. This system-bath coupling can be likewise handled by again performing a transformation \cite{c.c} given by
\begin{eqnarray}
q_{2} &=& Q_{2} \cos \theta + Q \sin \theta,
\label{q2}
\\
q &=& -Q_{2} \sin \theta + Q \cos \theta.
\label{q}
\end{eqnarray}
For simplicity, we assume that the new frequency $\Omega_{2}$ is the same with that of $\omega_{q}$ in eq. (\ref{l3new}). This allows us to rewrite eqs. (\ref{l2new}) and (\ref{l3new}) as 
\begin{eqnarray}
L_{2,3}&=&\frac{1}{2}m\dot{Q}_{2}^2 + \frac{1}{2}m\dot{Q}^2 + \frac{1}{2}m\Omega_{2}^2 Q_{2}^2 + \frac{1}{2}m\Omega_{2}^2 Q^2
\nonumber
\\
&+&\sqrt{2}C \cos \theta \sin \theta Q_{2}^2 - \sqrt{2}C \cos \theta \sin \theta Q^2
\nonumber
\\
&-&\sqrt{2}C\cos 2\theta Q_{2}Q.
\label{l23}
\end{eqnarray} 
Then regrouping and simplification yields
\begin{equation}
L_{2,3}=\frac{1}{2}m\dot{Q}_{2}^2 + \frac{1}{2}m\dot{Q}^2 + AQ_{2}^2-BQ^2-DQ_{2}Q,
\label{l23new}
\end{equation}
where
\begin{eqnarray}
A &=& \frac{1}{2}m\Omega_{2}^2 + \sqrt{2}C \cos \theta \sin \theta,
\label{A}
\\
B &=& \frac{1}{2}m\Omega_{2}^2 - \sqrt{2}C \cos \theta \sin \theta,
\label{B}
\\
D &=& \sqrt{2}C \cos 2\theta.
\label{D}
\end{eqnarray}
Likewise to eliminate the system-bath coupling, $D$ must again vanish, so that eq. (\ref{D}) yields
\begin{equation}
\theta = \frac{(2n+1)\pi}{4},
\label{Dsolution}
\end{equation}
where $n=0,1,2,...$. Imposing the condition of eq. (\ref{Dsolution}), we finally obtain a separable $L_{2,3}$ and thus the total Lagrangian components in eqs. (\ref{l1new}), (\ref{l2new}) and (\ref{l3new}) become
\begin{eqnarray}
L_{1} &=& \frac{1}{2}m\dot{q}_{1}^2 + \frac{1}{2}m\Omega_{1}^2 q_{1}^2,
\label{l1newest}
\\
L_{2} &=& \frac{1}{2}m\dot{Q}_{2}^2 + \frac{1}{2}m\Phi_{2}^2 Q_{2}^2,
\label{l2newest}
\\
L_{3} &=& \frac{1}{2}m\dot{Q}^2 + \frac{1}{2}m\Phi^2{Q}^2,
\label{l3newest}
\end{eqnarray}
where $\Phi_{2}^2=\Omega_{2}^2 + \frac{\sqrt{2}C}{m}$ and $\Phi^2=\Omega_{2}^2 - \frac{\sqrt{2}C}{m}$. Clearly, it is evident that the total Lagrangian is separable into propagators for three independent harmonic oscillators which enable us to smoothly evaluate the Feynman path integration in the context of white noise analysis. Moreover, the classical action can be written as $S = \int_{0}^{t}L_{1}d\tau + \int_{0}^{t}L_{2}d\tau + \int_{0}^{t}L_{3}d\tau \Rightarrow S_{1}+S_{2}+S_{3}$. Thus, the full propagator can be written as $K(q_{1},Q_{2},Q;q_{1o},Q_{2o},Q_{o};\tau) =K_{F}=K(q_{1},q_{1o};\tau) K(Q_{2},Q_{2o};\tau) K(Q_{2},Q_{o};\tau)$ where 
\begin{eqnarray}
K(q_{1},q_{1o};\tau)=K_{q_{1}}=\int_{}^{}\exp \left[ \frac{i}{\hbar}S_{1} \right] D[q_{1}],
\label{k1}
\\ 
K(Q_{2},Q_{2o};\tau)=K_{Q_{2}}=\int_{}^{}\exp \left[ \frac{i}{\hbar}S_{2} \right] D[Q_{2}],
\label{k2}
\\
K(Q,Q_{o};\tau)=K_{Q}=\int_{}^{}\exp \left[ \frac{i}{\hbar}S_{3} \right] D[Q].
\label{k3}
\end{eqnarray}
Hence, we have shown that the full propagator is likewise separable. We then proceed with the evaluation of each individual propagator using white noise analysis.

\subsection{The Evaluation of $K_{q_{1}}$}

We substitute eq. (\ref{l1newest}) and the classical action into eq. (\ref{whitenoisepropagator}) and in doing so, we obtain the following propagator:
\begin{eqnarray}
K_{q_{1}}&=&N \int \exp \left[ \frac{i+1}{2}\int_{0}^{t}\omega(\tau)^2 d\tau \right] 
\label{whitenoisepropagatorK1}
\\
&\times&\exp \left[ -\frac{i}{\hbar}\int_{0}^{t}S_{V}(q_{1})d\tau \right] \delta(q(t)-q_{1}) d\mu(\omega),
\nonumber
\end{eqnarray}
where $S_{V}(q_{1})$ is just a term for the effective action of the harmonic oscillator potential. We parametrize the Donsker-delta function in eq. (\ref{donsker}) as follows:
\begin{eqnarray}
\delta(q(t)-q_{1})&=&\frac{1}{2\pi}\int_{-\infty}^{+\infty}\exp \left[ i\lambda (q_{1o}-q_{1}) \right] 
\nonumber
\\
&\times&\exp \left[ i\lambda \int_{0}^{t}\omega(\tau)d\tau \right] d\lambda, 
\label{donsker2}
\end{eqnarray}
We also parametrize the second exponential expression, which contains the potential, at the right hand side of eq. (\ref{whitenoisepropagatorK1}), which yields: 
\begin{equation}
\exp \left[ -\frac{i}{\hbar}\int_{0}^{t}\frac{1}{2}m\Omega_{1}^2 \left(q_{1o}+\int_{0}^{t}\omega(\tau)d\tau \right)^2 d\tau \right].
\end{equation}
This contains second degree in white noise which makes it difficult to deal with. To remedy this, we apply Taylor series expansion as specified in ref. \cite{b.a}
\begin{eqnarray}
S_{V}(q_{1})&\approx& S_{V}(q_{1o}) + \frac{1}{1!}\int{}{}d\tau \omega(\tau) \frac{\partial S_{V}(q_{1o})}{\partial \omega(\tau)} 
\label{taylor}
\\
&+& \frac{1}{2!}\int{}{}d\tau_{1}d\tau_{2} \omega(\tau_{1}) \frac{\partial^2 S_{V}(q_{1o})}{\partial \omega(\tau_{1})\partial \omega(\tau_{2})}\omega(\tau_{1}). 
\nonumber
\end{eqnarray}
For simplicity we choose the initial point $q_{1o}=0$ which leads to $S_{V}(q_{1o})=0$ and
\begin{eqnarray}
S'&=&\frac{\partial S_{V}(0)}{\partial \omega(\tau)}=\frac{\hbar}{m}\int V'(0) d\tau \Rightarrow 0,  
\label{S'}
\\
S"&=&\frac{\partial^2 S_{V}(0)}{\partial \omega(\tau_{1})\partial \omega(\tau_{2})}=\frac{\hbar}{m}\int_{\tau_{1}\vee \tau_{2}}^{t}V''(0) d\tau 
\nonumber
\\
&\Rightarrow& \hbar \Omega_{1}^2(t-\tau_{1} \vee \tau_{2}).
\label{S"}
\end{eqnarray}
Then, with eqs. (\ref{donsker2}), (\ref{S'}) and (\ref{S"}) we can rewrite the propagator as
\begin{equation}
K_{q_{1}}=\int_{-\infty}^{+\infty}\frac{\exp \left[ -i\lambda q_{1} \right]}{2\pi}\left[ T.I\left(\sqrt{\frac{\hbar}{m}}\lambda \right) \right] d\lambda, 
\label{K1dim}
\end{equation}
where
\begin{equation}
I=N \exp \left[ -\frac{1}{2}\left\langle \omega, -(i+1)\omega \right\rangle \right] \exp \left[ -\frac{1}{2}\left\langle \omega, \frac{i}{\hbar}S"\omega \right\rangle \right],
\label{Iho}
\end{equation} 
and the evaluation of the Feynman path integral is carried by the T-transform \cite{b.a} given by $T.I(\xi=\sqrt{\frac{\hbar}{m}}\lambda) = \int_{}^{}I \exp \left[ i\left\langle \omega, \sqrt{\frac{\hbar}{m}}\lambda \right\rangle \right] d\mu(\omega)$ which can be simplified as
\begin{eqnarray}
T.I&=&\left[ det\left( 1+L(K+1)^{-1}\right)\right]^{-\frac{1}{2}}
\label{T-transform}
\\
&\times&\exp \left[ -\frac{1}{2}(K+L+1)^{-1}\int_{0}^{t}\left( \sqrt{\frac{\hbar}{m}}\lambda \right)^2 d\tau \right], 
\nonumber
\end{eqnarray}
where $K=-(i+1)$ and $L=i\hbar^{-1}S"$. Then, we substitute eq. (\ref{T-transform}) into eq. (\ref{K1dim}). Simplifying, we obtain
\begin{eqnarray}
K_{q_{1}}&=&\frac{1}{2\pi}\left[ det\left( 1+L(K+1)^{-1}\right)\right]^{-\frac{1}{2}}\int_{-\infty}^{+\infty}d\lambda 
\nonumber
\\
&\times&\exp \left[ \frac{-\hbar t(K+L+1)^{-1}}{2m}\lambda^2-iq_{1} \lambda \right].
\label{whitenoisepropagatorK1dim2}
\end{eqnarray}
However, we note that $\left( 1+L(K+1)^{-1}\right)=(1-\hbar^{-1}S")$ and $(K+L+1)^{-1}=i(1-\hbar^{-1}S")^{-1}$. We can then rewrite eq. (\ref{whitenoisepropagatorK1dim2}) as 
\begin{eqnarray}
K_{q_{1}}&=&\frac{1}{2\pi}\left[ det(1-\hbar^{-1}S")\right]^{-\frac{1}{2}}\int_{-\infty}^{+\infty}d\lambda 
\nonumber
\\
&\times&\exp \left[ \frac{-i\hbar t(1-\hbar^{-1}S")^{-1}}{2m}\lambda^2-iq_{1} \lambda \right].
\label{whitenoisepropagatorK1dim2change}
\end{eqnarray}
Utilizing the Gaussian integral formula, we obtain
\begin{eqnarray}
K_{q_{1}}&=&\frac{1}{2\pi}\left[ det(1-\hbar^{-1}S")\right]^{-\frac{1}{2}} 
\nonumber
\\
&\times&\sqrt{\frac{2\pi m}{i\hbar t \left\langle e, (1-\hbar^{-1}S")e \right\rangle}}
\nonumber
\\
&\times&\exp \left[ \frac{imq_{1}^{2}}{2\hbar t \left\langle e, (1-\hbar^{-1}S")e \right\rangle} \right],
\label{whitenoisepropagatorK1dim3}
\end{eqnarray}
where the unit vector $e$ is defined as $e=t^{-\frac{1}{2}}\chi_{[0,t]}$. Then, after some simplification \cite{b.a, c.d} we get 
\begin{eqnarray}
det(1-\hbar^{-1}S") &=& \cos \Omega_{1}t,
\label{determinant}
\\ 
\left\langle e, (1-\hbar^{-1}S")e \right\rangle &=& \frac{1}{\Omega_{1}t}\tan \Omega_{1}t.
\label{diagonalization}
\end{eqnarray}
Finally, using eqs. (\ref{determinant}) and (\ref{diagonalization}), we obtain the $q_{1}$-dimension propagator as
\begin{eqnarray}
K_{q_{1}}=\sqrt{\frac{m\Omega_{1}}{2\pi i\hbar t \sin \Omega_{1}t}} \exp \left[ \frac{im\Omega_{1}}{2\hbar}q_{1}^2\cot \Omega_{1}t \right].
\label{whitenoisepropagatorK1dimfinal}
\end{eqnarray}

\subsection{The Evaluation of $K_{Q_{2}}$ and $K_{Q}$ Propagators}

Notice that the Lagrangians $L_{2}$ and $L_{3}$ are just similar to that of the $q_{1}$-dimension Lagrangian. Thus by following the same procedure in evaluating $q_{1}$-dimension propagator, we obtain the propagators as
\begin{eqnarray}
K_{Q_{2}}=\sqrt{\frac{m\Phi_{2}}{2\pi i\hbar t \sin \Phi_{2}t}} \exp \left[ \frac{im\Phi_{2}}{2\hbar}Q_{2}^2\cot \Phi_{2}t \right],
\label{whitenoisepropagatorK2dimfinal}
\\
K_{Q}=\sqrt{\frac{m\Phi}{2\pi i\hbar t \sin \Phi t}} \exp \left[ \frac{im\Phi}{2\hbar}Q^2\cot \Phi t \right].
\label{whitenoisepropagatorK3dimfinal}
\end{eqnarray}

\subsection{The Full Propagator}
We can now solve for the full propagator which is just the product of eqs. (\ref{whitenoisepropagatorK1dimfinal}), (\ref{whitenoisepropagatorK2dimfinal}) and (\ref{whitenoisepropagatorK3dimfinal}). However, in doing so we must transform it back to its original coordinates. Doing so gives us the following relations:
\begin{eqnarray}
Q_{2} &=& q_{2} \cos \theta - q \sin \theta,
\label{Q2}
\\
Q &=& q_{2} \sin \theta + q \cos \theta.
\label{q}
\end{eqnarray}
Using these expressions for $Q$ and $Q_{2}$, with $\theta=\frac{(2n+1)\pi}{4}$, we can rewrite eqs. (\ref{whitenoisepropagatorK2dimfinal}) and (\ref{whitenoisepropagatorK3dimfinal}) as
\begin{eqnarray}
K_{Q_{2}}&=&\sqrt{\frac{m\Phi_{2}}{2\pi i\hbar t \sin \Phi_{2}t}}\exp \left[ \frac{im\Phi_{2}}{2\hbar}(q_{2}-q)^2\cot \Phi_{2}t \right],
\nonumber
\\
\label{whitenoisepropagatorK2dimfinalnew}
\\
K_{Q}&=&\sqrt{\frac{m\Phi}{2\pi i\hbar t \sin \Phi t}}\exp \left[ \frac{im\Phi}{2\hbar}(q_{2}+q)^2\cot \Phi t \right].
\label{whitenoisepropagatorK3dimfinalnew}
\end{eqnarray}
Finally transforming back $q_{1}, q_{2}$ into $x_{1}, x_{2}$ we obtain the relation
\begin{eqnarray}
q_{1} &=& x_{1} \cos \phi - x_{2} \sin \phi,
\label{q1x1}
\\
q_{2} &=& x_{1} \sin \phi + x_{2} \cos \phi,
\label{q2x2}
\end{eqnarray}
which allows us to rewrite eqs. (\ref{whitenoisepropagatorK1dimfinal}), (\ref{whitenoisepropagatorK2dimfinalnew}) and (\ref{whitenoisepropagatorK3dimfinalnew}) as
\begin{eqnarray}
K_{q_{1}}&=&\sqrt{\frac{m\Omega_{1}}{2\pi i\hbar t \sin \Omega_{1}t}}\exp \left[ \frac{im\Omega_{1}}{4\hbar}(x_{1}-x_{2})^2\cot \Omega_{1}t \right], 
\nonumber
\\
\label{whitenoisepropagatorK1dimfinal2}
\\
K_{Q_{2}}&=&\sqrt{\frac{m\Phi_{2}}{2\pi i\hbar t \sin \Phi_{2}t}}
\nonumber
\\
&\times&\exp \left[ \frac{im\Phi_{2}}{4\hbar}[\frac{\sqrt{2}}{2}(x_{1}+x_{2})-q]^2\cot \Phi_{2}t \right],
\label{whitenoisepropagatorK2dimfinalnew2}
\\
K_{Q}&=&\sqrt{\frac{m\Phi}{2\pi i\hbar t \sin \Phi t}}
\nonumber
\\
&\times&\exp \left[ \frac{im\Phi}{4\hbar}[\frac{\sqrt{2}}{2}(x_{1}+x_{2})+q]^2\cot \Phi t \right]. 
\label{whitenoisepropagatorK3dimfinalnew2}
\end{eqnarray}
Hence we can write the full propagator as 
\begin{eqnarray}
K_{F}&=&\left(\frac{m}{2\pi i\hbar t}\right)^{\frac{3}{2}}\left[ \frac{\Omega_{1}\Phi \Phi_{2}}{\sin \Omega_{1}t \sin \Phi t \sin \Phi_{2}t} \right]^{\frac{1}{2}} 
\nonumber
\\
&\times&\exp \left[ \frac{im\Omega_{1}}{4\hbar}(x_{1}-x_{2})^2\cot \Omega_{1}t \right]
\nonumber
\\
&\times&\exp \left[ \frac{im\Phi_{2}}{4\hbar}[\frac{\sqrt{2}}{2}(x_{1}+x_{2})-q]^2\cot \Phi_{2}t \right] 
\nonumber
\\
&\times&\exp \left[ \frac{im\Phi}{4\hbar}[\frac{\sqrt{2}}{2}(x_{1}+x_{2})+q]^2\cot \Phi t \right],
\label{kfull}
\end{eqnarray}
where the frequencies are given by $\Omega_{1}=\sqrt{\omega^2 + \frac{\lambda}{m}}$, $\Phi=\sqrt{\omega^2 - \frac{\lambda}{m}-\frac{\sqrt{2}C}{m}}$, and $\Phi_{2}=\sqrt{\omega^2 - \frac{\lambda}{m}+\frac{\sqrt{2}C}{m}}$. Now, it follows that the Liouville space propagator can be solved as $J_{F}=\left\langle K_{F}K_{F'}^*\right\rangle_{B}$, wherein in this operation, we trace out the bath variable $q$ and set the coupling constant $C$ to zero. In doing so, we find that $\Phi=\Phi_{2}$. We then obtain the evolution of the reduced density matrix in eq. (\ref{vonformal2}) as
\begin{eqnarray}
\rho_{S}(t)&=&\left(\frac{m}{2\pi \hbar t}\right)^{2}\left[ \frac{\Omega_{1}\Phi_{2}}{\sin \Omega_{1}t \sin \Phi_{2}t} \right] 
\nonumber
\\
&\times&\exp \left[ \frac{im\Omega_{1}}{4\hbar}[(x_{1}-x_{2})^2-(x'_{1}-x'_{2})^2]\cot \Omega_{1}t \right]
\nonumber
\\
&\times&\exp \left[ \frac{im\Phi_{2}}{4\hbar}[(x_{1}+x_{2})^2-(x'_{1}+x'_{2})^2]\cot \Phi_{2}t \right]
\nonumber
\\
&\times&\rho(0).  
\label{kfullmaster}
\end{eqnarray}  
We note that the obtained Liouville space propagator $J_{F}$ corresponds to the quantum mechanical propagator for coupled harmonic oscillators which agrees with the result obtained in ref. \cite{c.c}. 

\section{Conclusion}
In this article, we have successfully solved for the quantum Feynman propagator for a system of coupled harmonic oscillators interacting with a bath consisting of a single multimode harmonic oscillator using the white noise analysis. The full quantum propagator is a product of three harmonic oscillator propagators, which are obtained after imposing two successive coordinate transformations to decouple the oscillators from each other and from the bath. Furthermore, the obtained evolution of reduced density matrix corresponds to the evolution propagator of the coupled harmonic oscillators which has a form similar to ref. \cite{c.c}. 

Indeed, the method of white noise analysis posits promise in evaluating the propagators for open quantum systems. In particular, it can be applied to systems with N coupled oscillators which are all coupled to an environment, which can be used to model quantum transport of energy excitations in solid state and biological systems. The authors will explore these areas in further detail in future work.  

\acknowledgments
B. M. Butanas Jr would like to thank M. A. Pedroso-Butanas for the fruitful exchange of ideas and to the Philippines' Department of Science and Technology (DOST) and Central Mindanao University (CMU) for the scholarship and financial support via accelerated science and technology human resource development program (ASTHRDP) and a faculty development program (FDP), respectively. R. C. F. Caballar would like to thank M. A. A. Estrella for conceptual discussions that clarified matters related to this work. B. M. Butanas Jr. and R. C. F. Caballar would like to thank National Institute of Physics, College of Science and UP Diliman for support and for providing a stimulating research atmosphere.

\end{document}